\documentclass{conm-p-l}

\copyrightinfo{2011}{}

\usepackage{graphicx}
\usepackage{subfigure}

\usepackage{amssymb,amsmath}

\theoremstyle{definition}

\theoremstyle{remark}

\numberwithin{equation}{section}

\newcommand{\e}{\epsilon}

\newcommand{\Dl}{\Delta}

\newcommand{\al}{\alpha}
\newcommand{\be}{\beta}

\newcommand{\pa}{\partial}

\begin{document}

\title[On the paradox of pesticides]
{On the paradox of pesticides}

\author{Y. Charles Li}

\address{Department of Mathematics, University of Missouri, 
Columbia, MO 65211, USA}

\curraddr{}
\email{liyan@missouri.edu}
\urladdr{http://www.math.missouri.edu/~cli}

\author{Yipeng Yang} 

\address{Department of Mathematics, University of Missouri, 
Columbia, MO 65211, USA}

\email{yangyip@missouri.edu}

\thanks{}

\subjclass{Primary 92; Secondary 37}
\date{}

\dedicatory{}


\begin{abstract}
The paradox of pesticides was observed experimentally, which says that pesticides may dramatically 
increase the population of a pest when the pest has a natural predator. Here we use a mathematical 
model to study the paradox. We find that the timing for the application of pesticides is crucial 
for the resurgence or non-resurgence of the pests. In particular, regularly applying pesticides
is not a good idea as also observed in experiments \cite{HMCVJ07} \cite{Ham08}. In fact, the best 
time to apply pesticides is when the pest population is reasonably high.
\end{abstract}

\maketitle

\section{Introduction}

The paradox of pesticides says that pesticides may dramatically increase the population of a pest when 
the pest has a natural predator. Right after the application of the pesticide, of course the pest 
population shall decrease. But the pest may resurge later on resulting in a population well beyond 
the crop's economic threshold. 

In the experiment of \cite{LTH98}, pesticides (pyrethroids) are applied in apple orchards for the main goal of 
killing flying insects (lepidopteran and dipteran pests). The side effect of this is that later on the 
populations of other pests (phytophagous mites: panonychus ulmi and tetranychus urticae) feeding on the 
apple tree's leaves resurge above the crop's economic thresholds, while the populations of these pests on 
the leaves not sprayed with the pesticides remain low (below the crop's economic thresholds). These 
phytophagous mites are the preys of some predator mites (phytoseiid predator typhlodromus caudiglans and 
stigmaeid predator zetzellia mali). The pesticides kill both the leaf eating mites and the predator mites. 
One of the predator mites (typhlodromus caudiglans) is almost wiped out by the pesticides, and resurges very 
slowly. The population of the other predator mite (zetzellia mali) is only slightly affected by the 
pesticides, but it alone cannot control the resurgence of the leaf eating mites. The pesticide damage to 
the populations of the leaf eating mites is not as severe as that to the population of the predator mite 
(typhlodromus caudiglans). After the spraying of the pesticides, the dynamics described above lasts 11 weeks. 
The phenomena just described have been well documented in many experiments \cite{PCJ81} \cite{Har85} \cite{Har88}
\cite{GC89} \cite{Thi91} \cite{Har95} \cite{Mor98} \cite{JP02} \cite{Coh06}. 

In the experiment of \cite{HMCVJ07} \cite{Ham08}, pesticide effects on arthropods in the rice field are studied. 
There are four groups of arthropods: herbivores, detritivores, predators, parasitoids. Herbivores eat rice plants,
detritivores eat detritus in the rice field, and predators and parasitoids eat herbivores and detritivores. 
Pesticides can kill every group of arthropods. Since they move around faster and more often, predators contact the 
pesticides more often, and therefore are killed more. This leads to more abundance of herbivores. A new scheme 
for spraying pesticides was introduced in \cite{HMCVJ07}, that is, applying pesticides only when the pest densities 
are high enough, rather than applying regularly. In contrast to regularly applying pesticides every week, the new 
scheme dramatically reduces the density of herbivores among all arthropods, which are the main destroyers of the 
rice field; enhances the densities of predators and detritivores; and slightly reduces the density of parasitoids. 

Mathematical modeling on pest resurgence has also been studied, e.g. in \cite{MS08}, a time-discrete model 
generalizing the Nicholson-Bailey model \cite{NB35}, is studied and concludes that the paradox of pesticides 
could be caused essentially by the interspecific relationship and the intraspecific density effect. 

To avoid the paradox of pesticides, farmers can make use of Integrated Pest Management \cite{USE07} \cite{USE08}.
Mathematical models on integrated pest management are also studied \cite{Tan05}.

In this paper, we study a time-continuous model on pest resurgence, based upon the classical Lotka-Volterra 
model. We agree with the experimental work \cite{HMCVJ07} on that the timing of applying the pesticides is crucial 
for the paradox of pesticides. We believe that when the timing is right, even though the pesticides only kill the 
pests rather than their predators, the pests may still resurge in abundance. We also believe that the amount of 
pesticides applied is also important. We think that the issue of the amount of pesticides is more delicate than 
the theory of hormesis \cite{Luc68} \cite{Mor98} \cite{JP02} (that is, small amounts of pesticides enhance pest 
population, whereas large  amounts of pesticides reduce pest population).

\section{The mathematical model}

To model the effect of pesticides on pest resurgence, we study the Lotka-Volterra system with forcing,
\begin{eqnarray}
\frac{dH}{dt} &=& H(a-bP) - \al \Dl (t-T), \label{LVF1} \\
\frac{dP}{dt} &=& P(bcH-d) - \be \Dl (t-T), \label{LVF2}
\end{eqnarray}
where $H$ is the pest population, $P$ is the pest's predator population, ($a$, $b$, $c$, $d$, $\al$, $\be$) are 
positive constants, $\Dl (t)$ is an approximation of the delta function, and the $\Dl (t-T)$ terms represent 
the effects of pesticides. Specifically, we choose $\Dl (t-T)$ to be
\[
\Dl (t-T) = 1/\e , \text{ when } t \in [T, T+\e]; \ \ = 0, \text{ otherwise}.
\]

\begin{figure}[ht] 
\centering
\includegraphics[width=4.5in,height=4.5in]{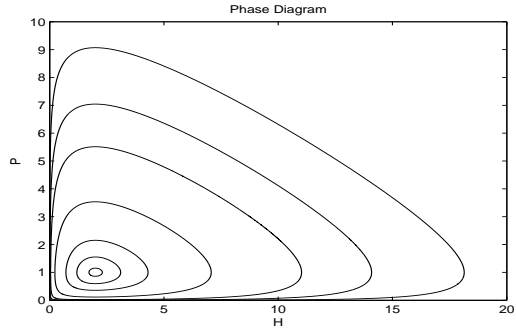}
\caption{The phase plane diagram of (\ref{LV1})-(\ref{LV2}), $a=b=d=1$, $c=0.5$, with different starting points 
of ($H, P$): ($1.8, 0.9$), ($1.4, 0.7$), ($1.0, 0.5$),($0.5, 0.25$), ($0.2, 0.1$), ($0.1, 0.05$), ($0.04, 0.02$).}
\label{FUF}
\end{figure}

Without the pesticide terms, the system (\ref{LVF1})-(\ref{LVF2}) reduces to the classical Lotka-Volterra system,
\begin{eqnarray}
\frac{dH}{dt} &=& H(a-bP) , \label{LV1} \\
\frac{dP}{dt} &=& P(bcH-d) , \label{LV2}
\end{eqnarray}
The nontrivial steady state of this system (\ref{LV1})-(\ref{LV2}),
\begin{equation}
H = \frac{d}{bc}, \quad P=\frac{a}{b}, \label{SS}
\end{equation}
is neutrally stable. The phase plane diagram of (\ref{LV1})-(\ref{LV2}) is shown in Figure \ref{FUF}.
\begin{figure}[ht] 
\centering
\includegraphics[width=4.5in,height=4.5in]{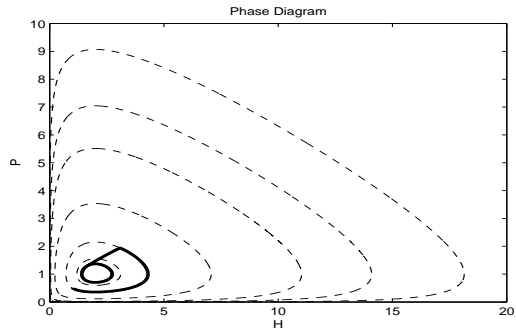}
\caption{A pesticide forced orbit of (\ref{LVF1})-(\ref{LVF2}), $a=b=d=1$, $c=0.5$, $T=3.5$, $\e = 0.07$, 
$\al = 1.4$, $\be = 0.7$. In this case, the pest population deceases and maintains at a lower amplitude oscillation.}
\label{FFD}
\end{figure}
\begin{figure}[ht] 
\centering
\includegraphics[width=4.5in,height=4.5in]{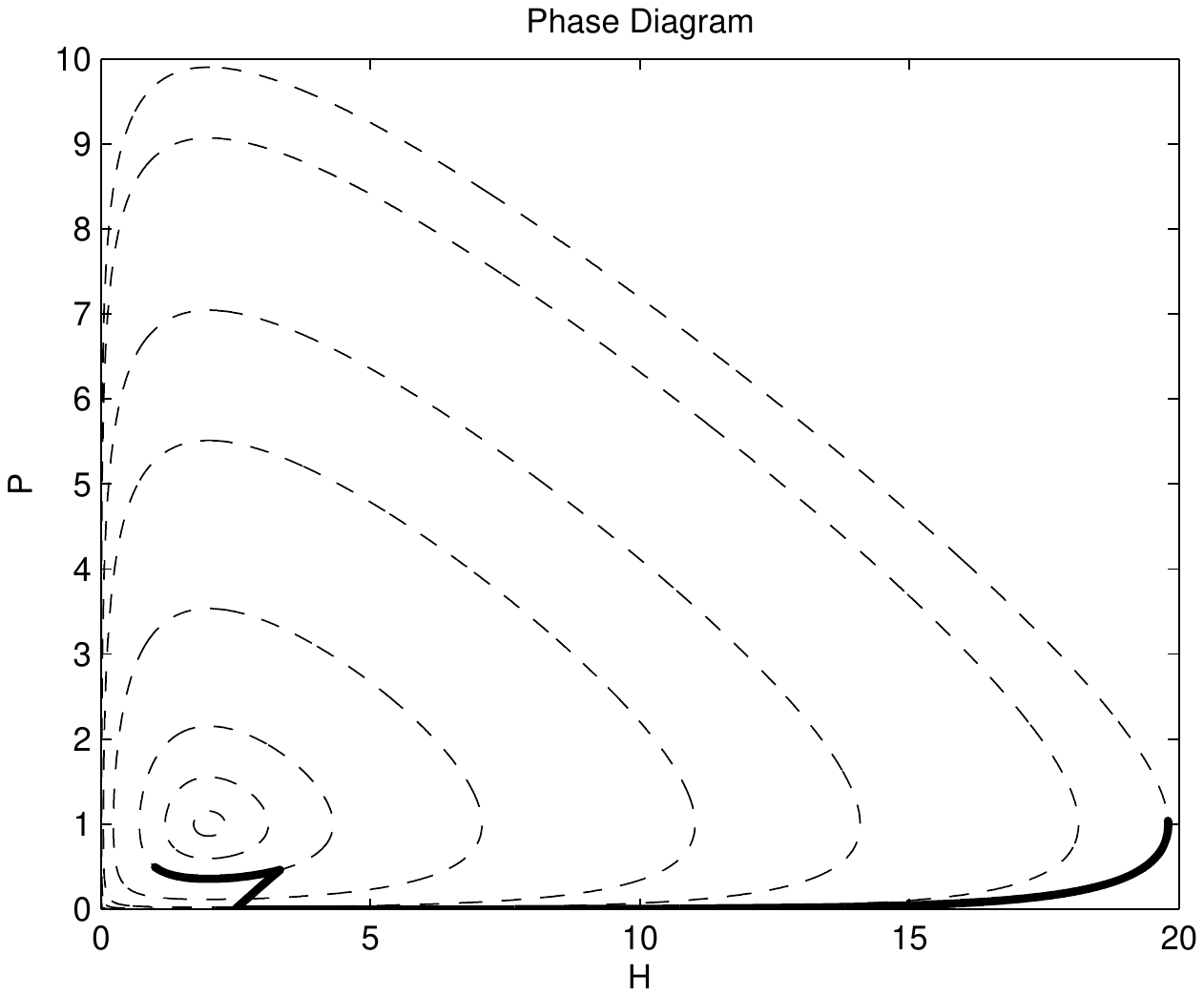}
\caption{A pesticide forced orbit of (\ref{LVF1})-(\ref{LVF2}), $a=b=d=1$, $c=0.5$, $T=2$, $\e = 0.047$, 
$\al = 0.94$, $\be = 0.47$. In this case, the pest population resurges with dramatic increase beyond the 
crop's economic threshold.}
\label{FFI1}
\end{figure}
\begin{figure}[ht] 
\centering
\includegraphics[width=4.5in,height=4.5in]{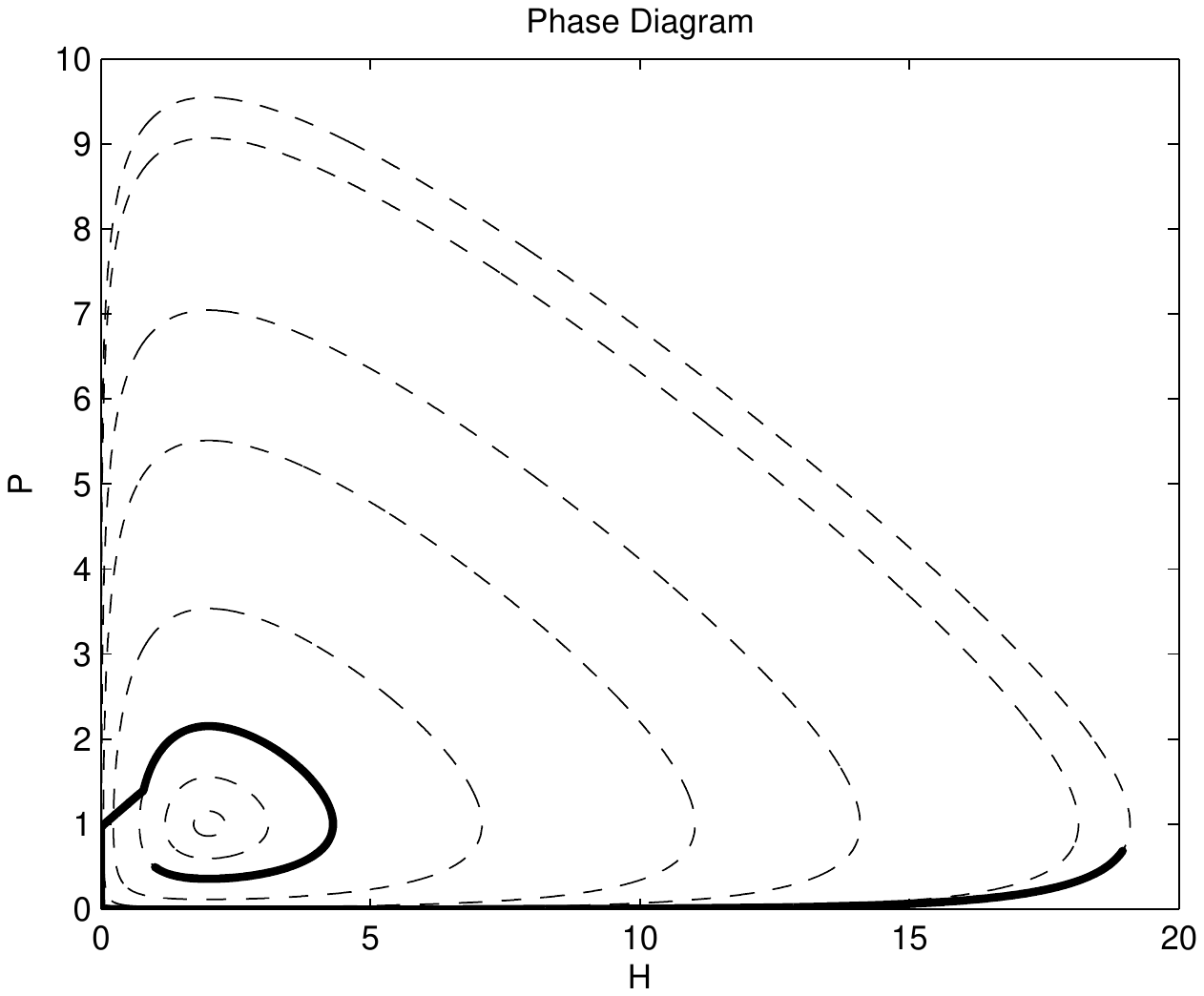}
\caption{A pesticide forced orbit of (\ref{LVF1})-(\ref{LVF2}), $a=b=d=1$, $c=0.5$, $T=5$, $\e = 0.039$, 
$\al = 0.78$, $\be = 0.39$. In this case, the pest population resurges with dramatic increase beyond the 
crop's economic threshold.}
\label{FFI2}
\end{figure}
\begin{figure}[ht] 
\centering
\includegraphics[width=4.5in,height=4.5in]{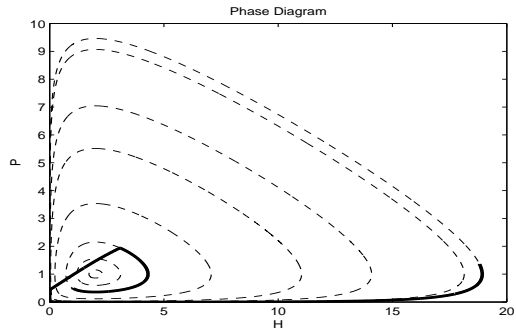}
\caption{A pesticide forced orbit of (\ref{LVF1})-(\ref{LVF2}), $a=b=d=1$, $c=0.5$, $T=3.5$, $\e = 0.149$, 
$\al = 2.98$, $\be = 1.49$. In contrast to Figure \ref{FFD}, here large amount of pesticides is applied,
and results in that the pest population resurges with dramatic increase beyond the 
crop's economic threshold.}
\label{FFI3}
\end{figure}

\begin{figure}[ht]
\centering
\subfigure[$Neumann \ b.c.$]{\includegraphics[width=2.3in,height=2.3in]{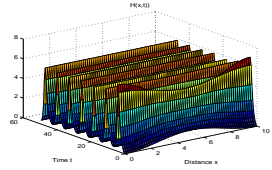}}
\subfigure[$Dirichlet \ initial$]{\includegraphics[width=2.3in,height=2.3in]{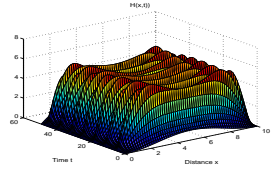}}
\subfigure[$Dirichlet \ transient$]{\includegraphics[width=2.3in,height=2.3in]{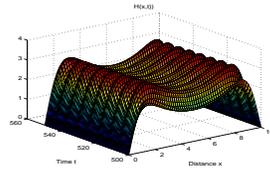}}
\subfigure[$Dirichlet \ asymptotic$]{\includegraphics[width=2.3in,height=2.3in]{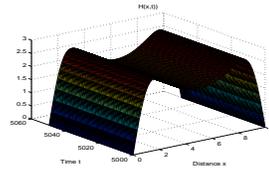}}
\caption{The asymptotic dynamics of the system (\ref{PDE1})-(\ref{PDE2}) without the pesticide terms ($\al = \be = 0$).
The specific initial condition used here is $H=P=\sin^2(\pi x/L)$. (a) is for the Neumann boundary condition.
(b)-(d) are for the Dirichlet boundary condition; (b) shows the dynamics of the initial temporal stage, (c) shows the transient stage dynamics, and (d) shows the asymptotic dynamics.}
\label{Apanel1}
\end{figure}

\begin{figure}[ht]
\centering
\subfigure[$Neumann \ b.c.$]{\includegraphics[width=2.3in,height=2.3in]{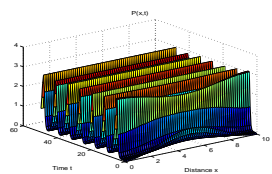}}
\subfigure[$Dirichlet \ initial$]{\includegraphics[width=2.3in,height=2.3in]{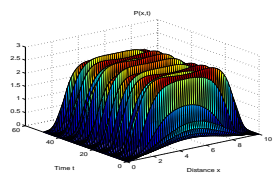}}
\subfigure[$Dirichlet \ transient$]{\includegraphics[width=2.3in,height=2.3in]{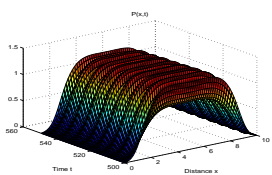}}
\subfigure[$Dirichlet \ asymptotic$]{\includegraphics[width=2.3in,height=2.3in]{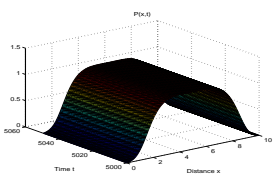}}
\caption{The same with Figure \ref{Apanel1} except for variable $P$.}
\label{Apanel2}
\end{figure}

\begin{figure}[ht] 
\centering
\includegraphics[width=4.5in,height=4.5in]{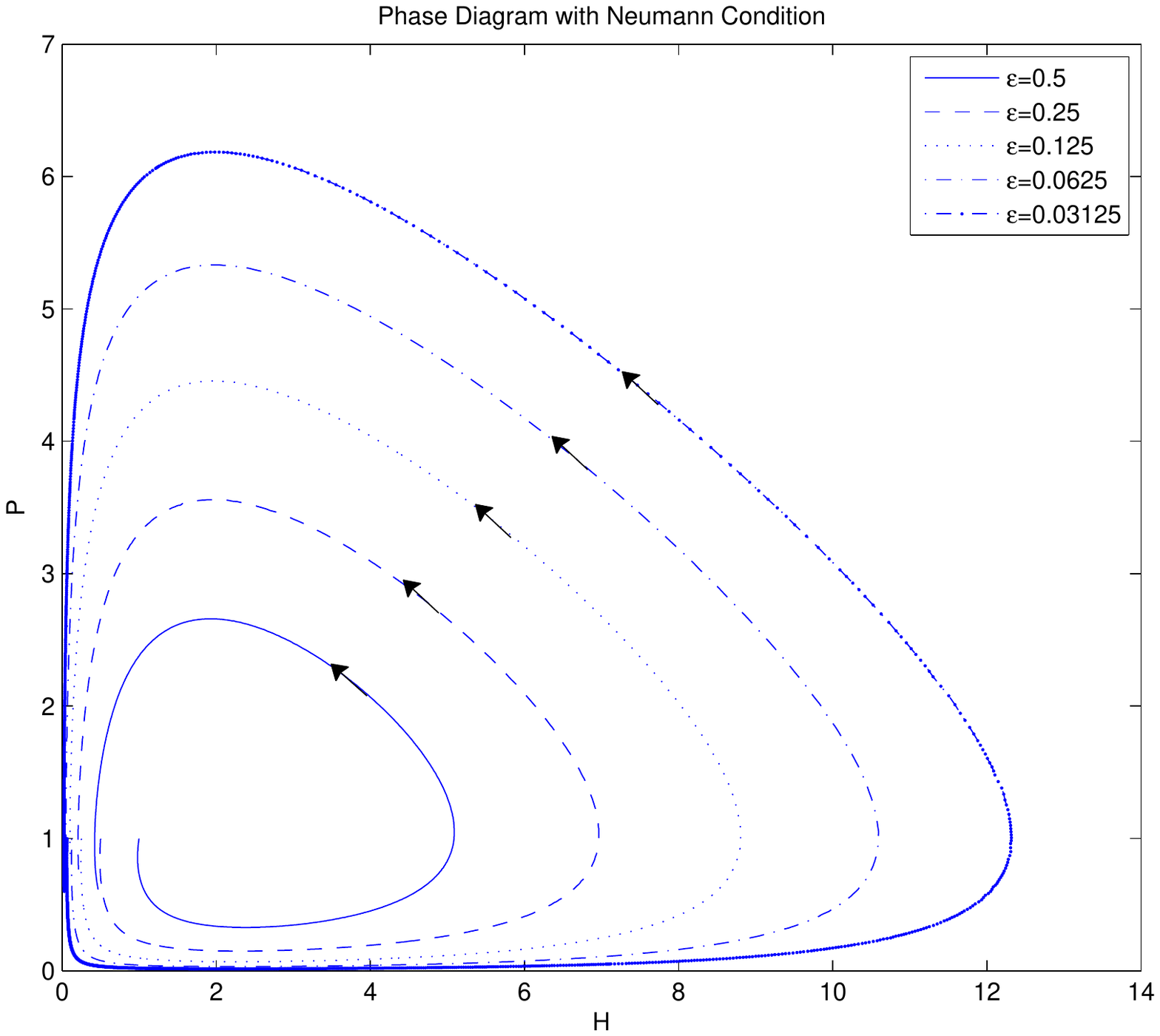}
\caption{The phase plane diagram of the spatial mid-points of $H(t,x)$ and $P(t,x)$ under 
the system (\ref{PDE1})-(\ref{PDE2}) without the pesticide terms ($\al = \be = 0$); and 
for the Neumann boundary condition with the intial condition $H = \e [1-cos(2\pi x/L)]$, $P = 1-cos(2\pi x/L)$
with various values of $\e$.} 
\label{F-NHP}
\end{figure}

\begin{figure}[ht]
\centering
\subfigure[$H \ profile$]{\includegraphics[width=2.3in,height=2.3in]{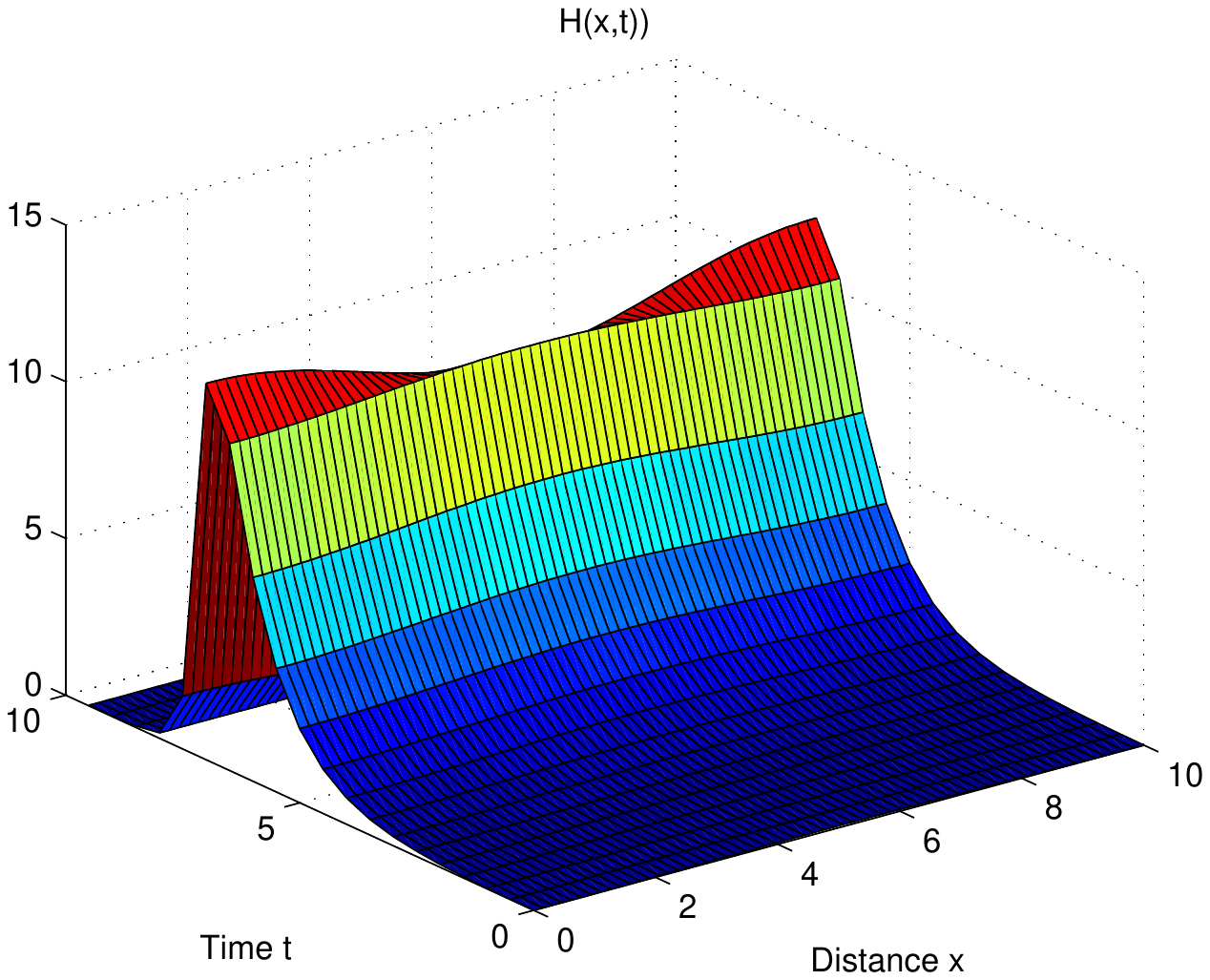}}
\subfigure[$P \ profile$]{\includegraphics[width=2.3in,height=2.3in]{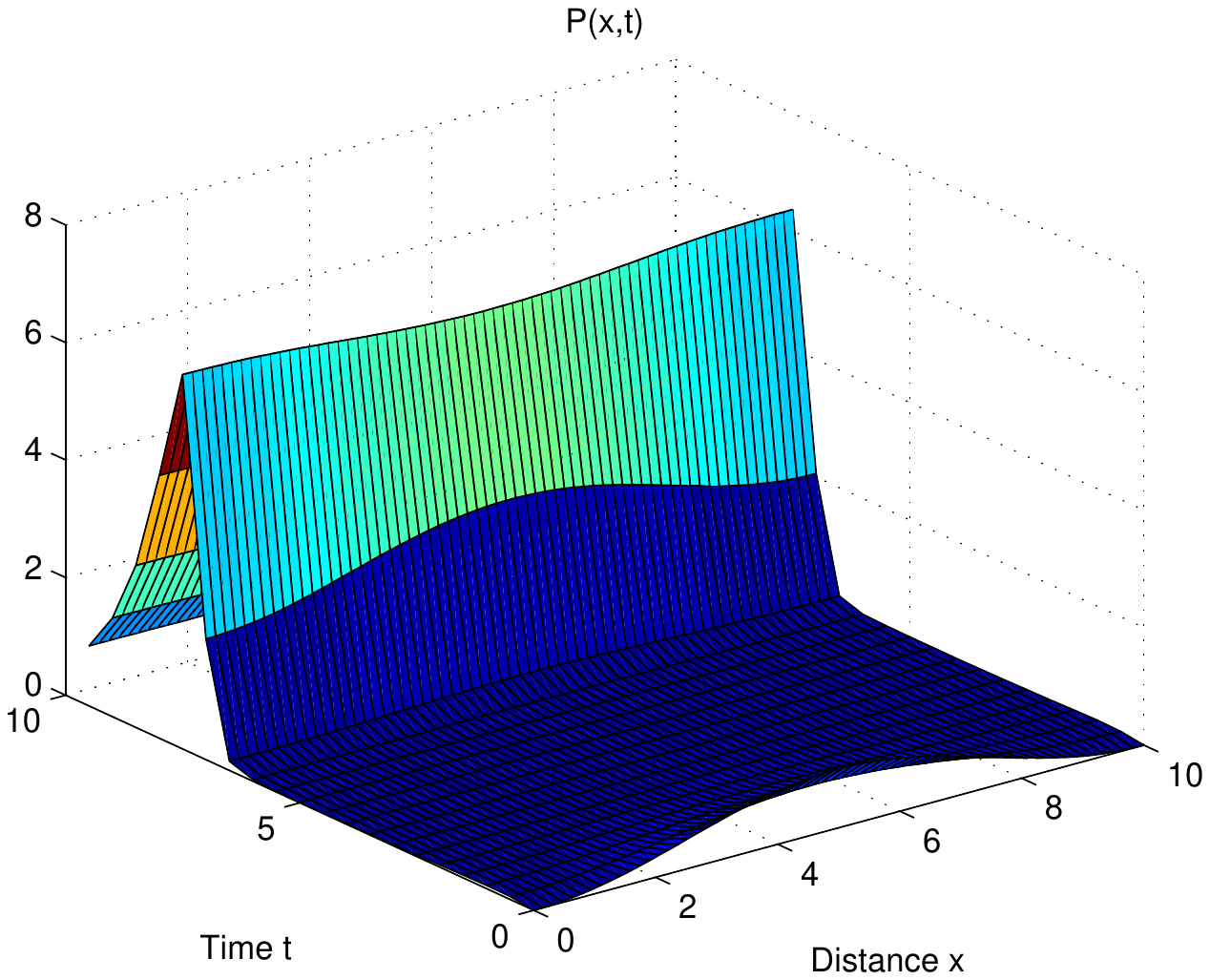}}
\caption{The the spatial-temporal profiles of $H$ and $P$ corresponding 
to the ($\e = 0.03125$) one-loop in Figure \ref{F-NHP}.}
\label{NHNP}
\end{figure}

\begin{figure}[ht] 
\centering
\includegraphics[width=4.5in,height=4.5in]{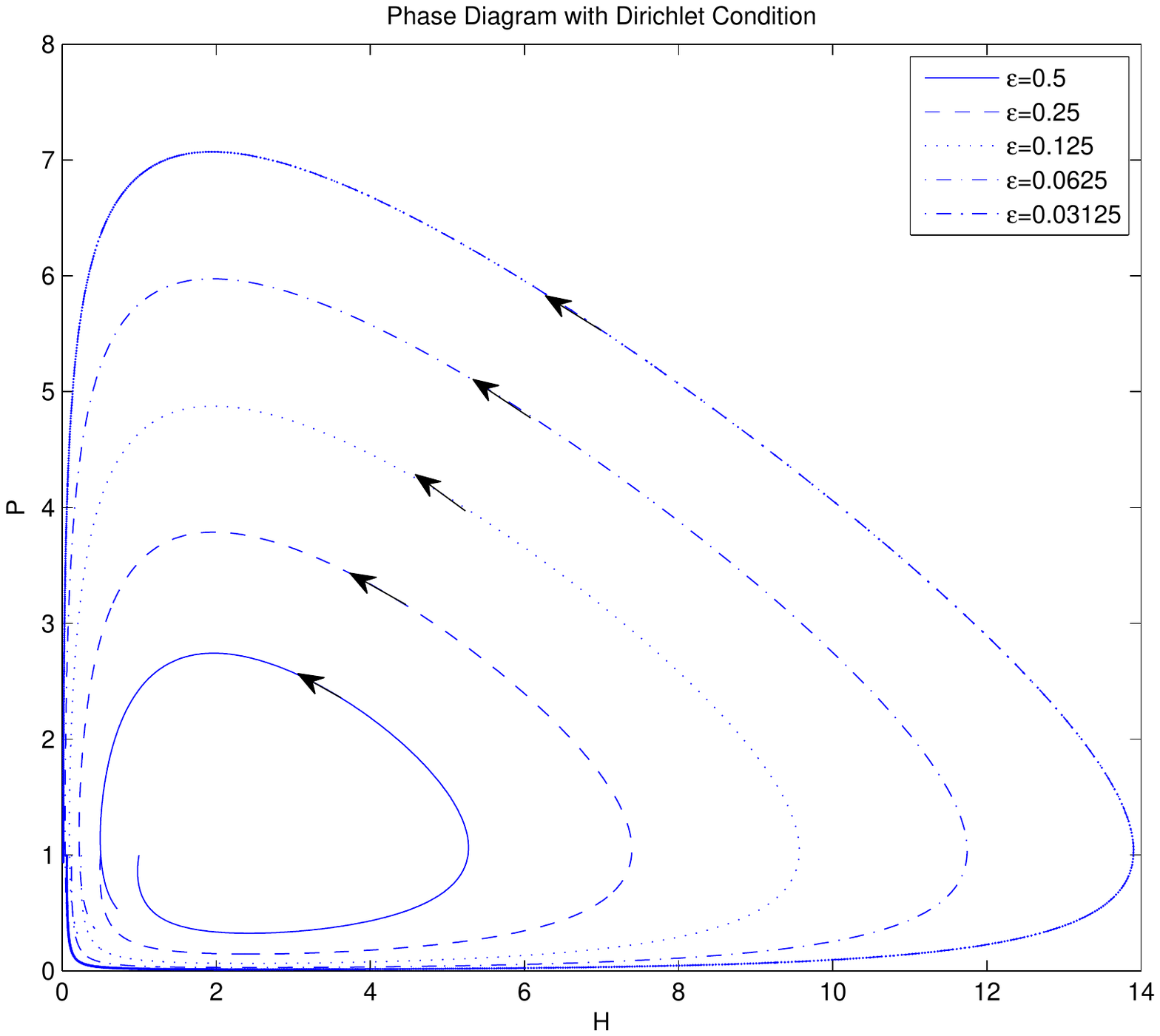}
\caption{The phase plane diagram of the spatial mid-points of $H(t,x)$ and $P(t,x)$ under 
the system (\ref{PDE1})-(\ref{PDE2}) without the pesticide terms ($\al = \be = 0$); and 
for the Dirichlet boundary condition with the intial condition $H = \e [1-cos(2\pi x/L)]$, $P = 1-cos(2\pi x/L)$
with various values of $\e$.} 
\label{F-DHP}
\end{figure}

\begin{figure}[ht]
\centering
\subfigure[$H \ profile$]{\includegraphics[width=2.3in,height=2.3in]{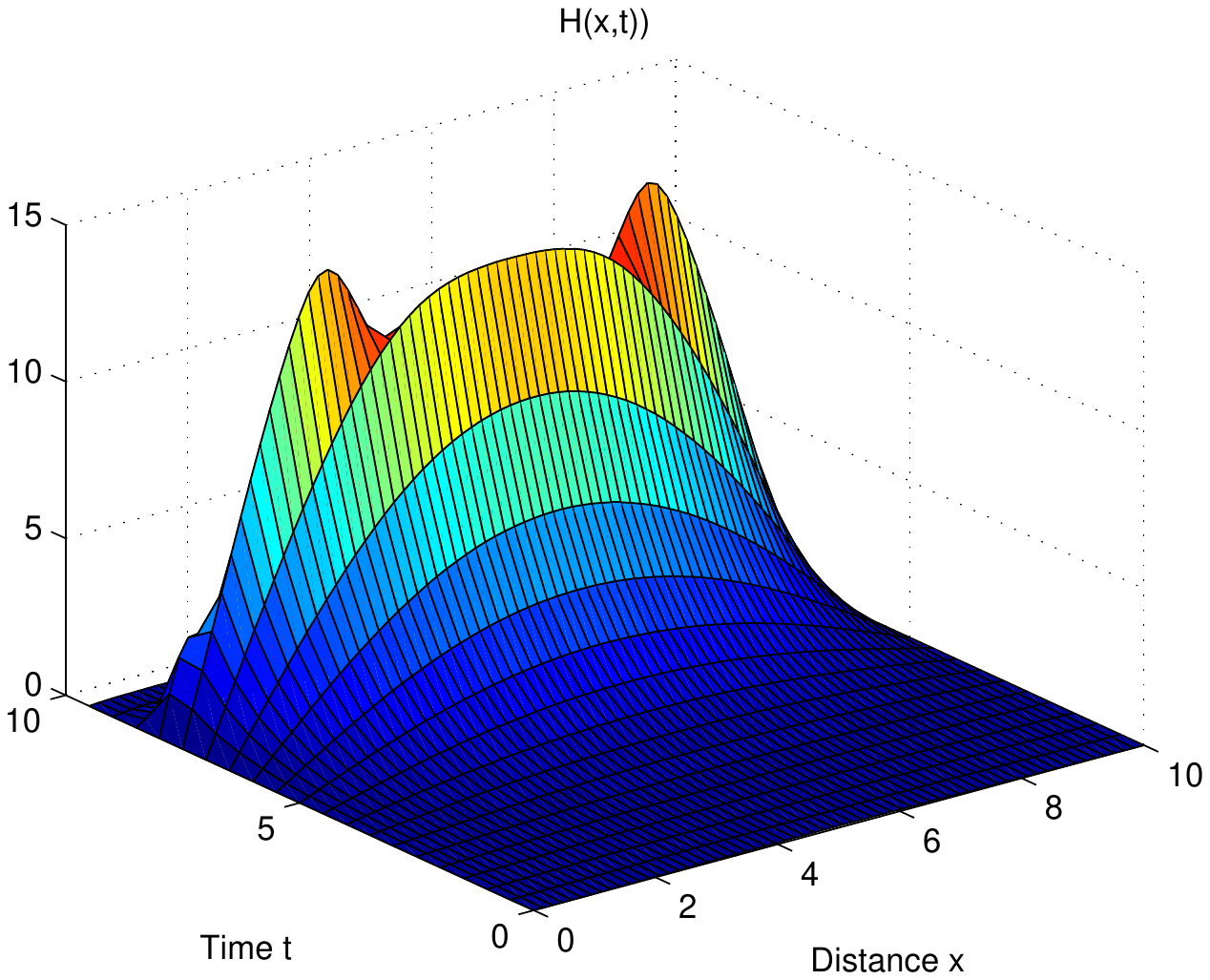}}
\subfigure[$P \ profile$]{\includegraphics[width=2.3in,height=2.3in]{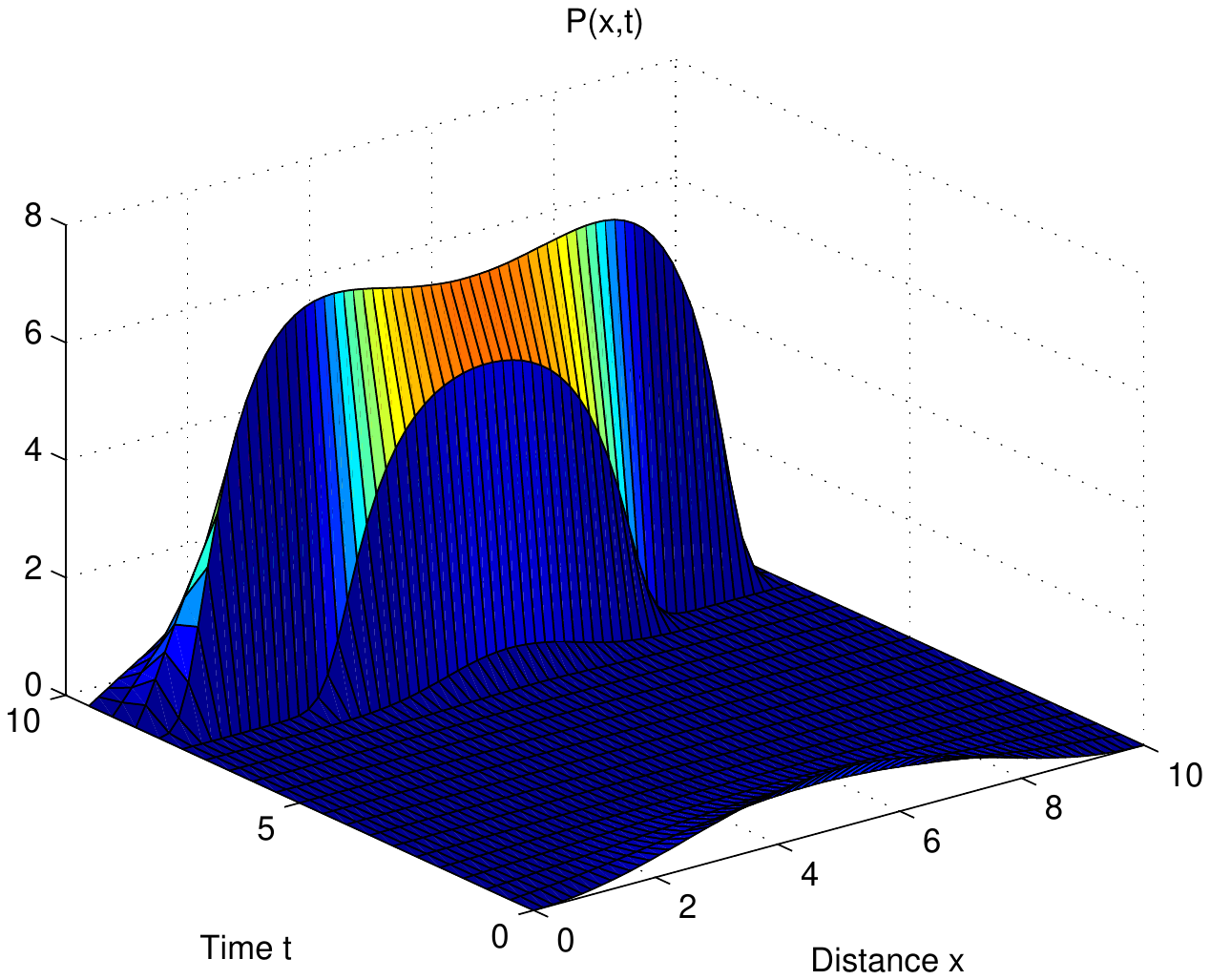}}
\caption{The the spatial-temporal profiles of $H$ and $P$ corresponding 
to the ($\e = 0.03125$) one-loop in Figure \ref{F-DHP}.}
\label{DHDP}
\end{figure}

The key points on the resolution of the paradox of pesticides are as follows:
\begin{enumerate}
\item The timing of applying the pesticides is crucial. If the pesticides are applied when the populations of both the 
pests and the predators are relatively large, then a decrease in both populations can be achieved, see Figure 
\ref{FFD}. On the other hand, if the pesticides are applied when either the pest's population or the 
predator's population is relatively small, then a dramatic increase in the resurgent pest's population occurs, leading 
to pest's population well beyond the crop's economic threshold, see Figures \ref{FFI1} and \ref{FFI2}. From these 
figures, it is clear that even though the pesticides only kill the pests rather than their predators (that is, 
right after the application of the pesticides, the pest's population decreases, while the predator's population 
maintains the same), the pests still resurge in abundance beyond the crop's economic threshold. This is because
that when the population of the pests decreases, the predator's population will decreases too, since the 
predators feed on the pests. It is the relatively minimal values of both the pest's population and the 
predator's population that decide how large the cycle which they are going to sit on in the phase plane.
\item The amount of pesticides applied is also important. In the case of Figure \ref{FFD}, if the amount 
of pesticides is large enough, a dramatic increase in the resurgent pest's population can still occur as 
shown in Figure \ref{FFI3}.
\end{enumerate}

\section{Spatial Effect}

To study the spatial effect, we introduce the following specific model,
\begin{eqnarray}
\frac{\pa H}{\pa t} - \frac{\pa P}{\pa x} \frac{\pa H}{\pa x} &=&\frac{\pa^2 H}{\pa x^2} + H(1-P) - 
\al \Dl (t-T, x), \label{PDE1} \\
\frac{\pa P}{\pa t} + \frac{\pa H}{\pa x} \frac{\pa P}{\pa x} &=& \frac{\pa^2 P}{\pa x^2} +  P(0.5 H-1) 
- \be \Dl (t-T, x), \label{PDE2}
\end{eqnarray}
where the second term on the left hand side represents 'convection' due to predator chasing and prey 
escaping, the first term on the right hand side represents population 'diffusion', and the last term on the 
right hand side represents pesticide effect now depending upon both space and time. Temporally the pesticide 
term is still near Delta function. Two type of boundary conditions are studied:
\begin{itemize}
\item Neumann boundary condition,
\[
\pa_x H = \pa_x P = 0, \ \ x= 0, L;
\]
\item Dirichlet boundary condition,
\[
H = P = 0, \ \ x= 0, L.
\]
\end{itemize}
Without the pesticide terms ($\al = \be = 0$), for a variety of initial conditions, we observe that the 
asymptotic state under the Neumann boundary condition is spatially uniform, while the 
asymptotic state under the Dirichlet boundary condition is the same steady state which appears to be an
attractor; see Figures \ref{Apanel1} and \ref{Apanel2}. 

For the paradox of pesticides, the asymptotic states are not important. One has to look at the entire temporal 
dynamics, in fact, the intial stage dynamics is much more important. Imagine that at the initial temporal stage,
the population of the pest reaches beyond the crop's economic threshold, the damage to the crops is done, then the 
later dynamics is meaningless. In Figure \ref{F-NHP}, we plot the `phase plane diagram' of the initial temporal stage dynamics of the spatial mid-points of $H(t,x)$ and $P(t,x)$ for the intial condition
\[
H = \e [1-cos(2\pi x/L)], \ P = 1-cos(2\pi x/L);
\]
with various values of $\e$. In Figure \ref{F-NHP}, we only show the most important initial `one-loop' for 
each orbit. This initial `one-loop' is the most relevant for the crop application. We view Figure \ref{F-NHP}
as the counterpart of Figure \ref{FUF}, and study the paradox of pesticide for the system 
(\ref{PDE1})-(\ref{PDE2}). In Figure \ref{NHNP}, the spatial-temporal profiles of $H$ and $P$ corresponding 
to the ($\e = 0.03125$) one-loop in \ref{F-NHP} are shown. The counterparts for the Dirichlet boundary condition
are shown in Figures \ref{F-NHP} and \ref{DHDP}.

\section{Conclusion}

Via a mathematical model, we studied the experimentally observed paradox of pesticides.
Our conclusion is that the timing for the application of pesticides is fundamental for the control of pest 
population. Regularly applying pesticides is not a good control as also observed in experiments \cite{HMCVJ07} 
\cite{Ham08}. In fact, the best time to apply pesticides is when the pest population is reasonably high.
We also studied a generalized partial differential equation model, we arrive at the same conclusion.

\end{document}